\documentstyle[epsf,twocolumn,aps]{revtex}
\begin{document}

\draft

\title{Exploration of resonances by analytic continuation in the coupling
constant}

\author{N. Tanaka$^1$, Y. Suzuki$^2$, and K. Varga$^{2,3}$}
\address{
$^1$Graduate School of Science and Technology, Niigata University,
Niigata 950-21, Japan\\
$^2$Department of Physics, Niigata University,
Niigata 950-21, Japan\\
$^3$Institute of Nuclear Research, Hungarian Academy of Sciences,
(ATOMKI),
H-4001 Debrecen, Hungary}

\date{\today}

\maketitle

\begin{abstract}
The energy and the width of  resonance states are determined
by analytic continuation of bound-state energies as a function of 
the coupling constant (potential strength). The advantage of the
method is that the existing techniques for calculation of  bound states 
can be applied, without any modifications, to determine the position of
resonances. Various numerical examples show the applicability of the 
method for three-body systems, including the excited states of the 
$^6$He and $^6$Li. 
\end{abstract}


\narrowtext
Although many of  excited states are resonances \cite{kb} in nuclear 
physics, they 
seldom receive special attention in theoretical nuclear structure 
calculations. The reason is obvious: The solution of the nuclear many-body
Schr\"odinger equation poses insurmountable difficulties even for bound 
states in most of the cases, and the solution for the resonances is mostly 
limited to two-body problems. The theoretical nuclear model calculations 
therefore usually treat the resonances as bound states. This approximation,
however, can only be justified for very narrow resonances.
\par\indent
In the recent years many theoretical and  experimental nuclear
physicists have concentrated their research activity on unstable nuclei
and the new phenomena discovered have revived the interest in
resonance states as well. These nuclei are very rich in resonance
states, and one of the most spectacular phenomena, the ``Borromean''
binding \cite{NBR}, is also closely related to resonances. In a 
Borromean system
(e.g., $^6$He or $^{11}$Li) the bound three-body system has no bound 
binary subsystem, but the two-body subsystems usually have a resonance.
Besides these two-body resonances, one can find an example for three-body 
resonances and for even more complicated ones as well 
\cite{NBR,matsui,afnan,csoto}.
\par\indent
Although there are elegant and powerful techniques to find resonances
as poles of the S-matrix (see e.g. \cite{kb}), these are mostly limited 
to simple interactions and systems. 
For composite nuclear systems the theoretical apparatus has
been formulated for finding the energies of bound states. The easiest way is 
thus to use the machinery developed for bound states to gain information
on resonances. There are indeed two methods widely used in nuclear
and atomic physics which rely on bound state type solutions, that is, use
square integrable functions. The first,
the ``real stabilization method'' \cite{HT}, solves the Schr\"odinger equation
in a box and exploits the fact that the energy of the resonance state
remains stable against the change of box size within certain limits
(the wave function is localized inside the box), while
the energies of the continuum states rapidly change.  The second, the
complex scaling method (CSM) \cite{CS}, rotates the coordinate ${\bf r}$ by 
$e^{i \theta}{\bf r}$ and transforms the continuum resonance wave 
function to the normalizable wave function of the bound state.
Both of these methods have the advantage that they use square integrable 
functions and therefore, after some modifications, the already existing 
methods of bound-state type can be applied. 
These two methods are known to be in an intimate connection \cite{t1}.
A slight disadvantage of the real stabilization method is that in order 
to determine 
the width of a resonances one has to solve the Schr\"odinger equation with 
many different box sizes and that it becomes computer time consuming for
composite systems. Moreover, the box for a many-particle system is a
many-dimensional object and sometimes it is difficult to find the 
appropriate intervals to be changed. In practical cases the method turns
out to be suitable for narrow resonances only. The complex scaling method
has proved to be capable of exploring wider resonances. 
Its computational burden is the calculation of complex matrix elements 
and the solution of complex eigenvalue problem. 
The source of the real hardship is that the Hamiltonian
is nonhermitian and the variational principle to find an energy
minimum cannot be applied. A  useful guiding principle 
for selection of the  basis, therefore, has been lost and it becomes 
complicated to choose the appropriate trial functions.
\par\indent
Kukulin {\it et.al.} \cite{accc} suggested an even simpler 
method to study resonant states; the analytical continuation in the 
coupling constant (ACCC). 
Their approach is based on the intuition that the resonances can
be thought as the continuation of bound states when the attraction of the 
interaction decreases. More precisely, they analytically continue the energy of the 
bound state  as a function of the strength of the potential to the 
complex plane to reveal the width and energy of the resonance.
The authors parameterize the square root of the energy by a 
Pad\'e approximation as a function 
of the coupling constant of the potential (potential strength). The
coefficients of the Pad\'e approximation are determined by solving the 
Schr\"odinger equation for the coupling constants giving bound states. 
The computational demand thus amounts to solving the
Schr\"odinger equation for bound states for several different potential
strengths. The Pad\'e approximation is chosen to approximate 
the square root of the energy as a function of the coupling constant 
because it is more general and more powerful
than other possibilities. For example, unlike a Taylor expansion it can
simulate singularities of a function near the threshold. 
\par\indent
The ACCC method was hitherto applied in a few simple test cases 
for two-body resonances e.g., the analytically solvable square-well 
potential and a macroscopic $\alpha$-$\alpha$ system with a simple potential.
The application was limited because to fix the coefficients of the
Pad\'e approximation one has to solve the bound state problem accurately.
Nowadays, there are several reliable methods at hand to solve few-body
problems. 
Before going to some really challenging applications, it is unavoidable
to test the performance and to learn the limitations of the ACCC method.
\par\indent
The aim of this paper is to show that the ACCC is really a 
powerful method to study the resonances of nuclear systems. 
A mathematical proof is not available, and the analytical solution 
is available only for a few very special cases. Therefore we must rely
on numerical examples for various cases. We compare the results of ACCC
to other methods such as direct numerical integration(DNI) or CSM.
\par\indent
The ACCC assumes that the Hamiltonian of the system is written as 
$ H(\lambda)=H_1+\lambda H_2 $ ,
where $H_2$ is the attractive part of the interaction. By decreasing $\lambda$
the bound state approaches the threshold (at
$\lambda_0$ the energy is $E(\lambda_0)=0$) and may become a resonance or
a virtual state. In Ref. \cite{accc} it has been shown that for a two-body
system, near the threshold
the square root of the energy behaves as $k_l(\lambda)\sim \sqrt{\lambda
-\lambda_0}$ for $l>0$ and $k_0(\lambda)\sim (\lambda-\lambda_0)$ 
for $S$-wave.
Introducing a variable $x=\sqrt{\lambda-\lambda_0}$
the analytic function $k_l$ has two branches $k_l(x)$ and $k_l(-x)$.
Due to the analyticity of these functions we can continue them
into the resonance region ($\lambda < \lambda_0$) from the bound states
$(\lambda>\lambda_0)$. Motivated by the above functional form of 
$k_l$ near the threshold, the Pad\'e approximation of the form \cite{accc}
\begin{equation}
\label{eqn:2}
k_l(x)=i \ {c_0+c_1 x+c_2x^2+\ldots +c_Mx^M\over
1+d_1 x+d_2 x^2+\ldots d_N x^N}
\end{equation}
is used for the analytical continuation.
In principle $c_0$ is to be zero but reserved to take care of possible 
errors in the determination of $\lambda_0$. In practice $c_0$ is found to 
be much smaller than other $c$ values.
\par\indent
The coefficients of the polynomials are calculated in the bound state
region and therefore are real. If $\lambda < \lambda_0$ $x$ is pure
imaginary and $k_l(x)$ becomes complex. The energy and width of the resonance
state is given by
$ E-i \Gamma/2={k_{l}}^{2} $ 
To determine the coefficients we are to solve the bound state problem for
various coupling constants $\lambda$ $(>\lambda_0)$ and try to
find the threshold value $\lambda_0$ $(k_l(\lambda_0)=0)$. To have a 
reliable approximation one has to know the accurate values of the 
coefficients in the Pad\'e approximation, that is, one has to solve
the bound state problem to high (typically 4 or more digits) accuracy, 
especially at the threshold. The stochastic variational approach 
\cite{Kuk1,svm} seems to be quite appropriate for this purpose. 
%%%%%%%%%%%%%%
\par\indent
Since one has to solve the bound state problem many times to determine 
the coefficients, one may think that the ACCC would be computer time 
consuming for larger systems. Due to the simple linear dependence of the
Hamiltonian on the coupling constant, however, one does not have to recalculate the matrix elements 
and the computational load is just the rediagonalization of the Hamiltonian
for different values of the coupling constant. This property makes the 
application especially simple.
\par\indent
To solve the bound state problem we use the stochastic variational method.
Of course any other method may be suitable. A point to be emphasized here
is that by changing the coupling constant one goes from deeply bound to 
weakly bound states and finally to the threshold. All these states should 
be accurately treated and in a variational calculation, for example, 
one has to choose a basis which adequately spans the configuration space.
%%%%%%%%%%%%%%
\par\indent
To illustrate the method we start with a simple example. We consider a
system of two-particles with mass $m$ interacting via 
a two-range Gaussian potential:\cite{cgkp}
\begin{equation}
\label{eqn:4}
V(r)=-8\lambda \ \exp[-(r/2.5)^{2}]+2 \ \exp[-(r/5)^{2}],
\end{equation}
where $\hbar=c=m=1$.
This simple problem can be easily solved by DNI \cite{gamow}. The trajectories of resonances of different
partial waves by ACCC and by DNI
are compared  in Fig. 1. For simplicity we assumed $M=N$ in Eq. (\ref{eqn:2}).
To determine the $2M+1$ coefficients one has to solve the bound state
for $2M+1$ different coupling constants $(\lambda_1,\ldots ,\lambda_{2M+1})$. Once the bound state energies are known, the coefficients of the 
Pad\'e approximation can be extracted by solving a system of linear equations. 
Due care must be taken to avoid numerical problems. 
The ACCC results agree very nicely with the DNI results 
including the $1S$ excited resonance state.
\par\indent
The second example tests the effectiveness of the method for wide resonances. 
The wide resonances may cause serious difficulties in many cases.
In this example we have used the same potential as Eq.(\ref{eqn:4}) 
but with a lower barrier ($V_{0} = 0.25$ instead of $V_{0} = 2$). 
The results of ACCC and DNI agree very well (see Fig. 2). 
The results are surprisingly good considering the fact that, 
in the last few points where the attraction is 
very weak, the widths of the states are two times of their energies (one
may not really call them resonances).
%%%%%%%%%%%%%%
\par\indent
The third example is a three-body case. Three bosons of mass, 
$\hbar^2/m = 41.47$, interact via the potential 
\begin{equation}
\label{eqn:5}
V(r)=-120 \ \exp[-r^{2}]+3 \lambda \ \exp[-(r/3)^{2}].
\end{equation}
The energy and length are in units of MeV and fm. 
Note that  there is no two-body bound state.
We compare the results of ACCC to those of CSM. This simple
example is selected because the CSM might have some inaccuracy in more
complex cases, and we want to address the applicability of ACCC
in a clean test case. The three-body bound state problem for 
zero total angular momentum has been solved by the stochastic 
variational method by using a Gaussian basis \cite{svm}. 
The trajectories of ACCC and CSM as a function of $\lambda$ are shown in
Fig. 3. The two methods give the same resonance position for a wide
range of potential strength, justifying the suitability of ACCC.
Pad\'e approximations with $M$=3,\ 5 and 7 are used.
By increasing the number of terms
the agreement slightly improves but already $M$=3 terms give good
results. Except for the threshold value, one has the freedom to
choose those coupling constants for which the bound state problem is
solved to determine the coefficients of the Pad\'e approximations.
The position of the resonance, of course, depends on the choice of
the set of $\lambda$. We found that if the values of coupling constants
used are distributed into a wide range the dependence of the
resonance parameters $E$ and $\Gamma$ on the input values are relatively
small. The sensitivity can be easily controlled by comparing the results 
starting from several adequately chosen input sets. 
\par\indent
The resonance states of the  $\alpha+N+N$ three-body model are chosen
as a practical example. These resonances have been 
very intensively studied in the past  \cite{matsui,afnan}
and the investigation is  continued with great elan \cite{NBR,csoto}.
Previously, we described the bound states of the $^6$He and $^6$Li
nuclei in an $\alpha+n+n$ and $\alpha+p+n$-type microscopic cluster model
with the stochastic variational method \cite{svm,asv}. To determine the 
resonances of these systems by ACCC, we used the same type of bound state 
solutions. 
The model assumes an alpha cluster but uses a fully microscopic six-body
wave function. The nucleons interact via the Minessota effective interaction
(sum of the Coulomb and spin-isospin dependent central and spin-orbit 
potentials) \cite{minessota}. The strength of the attraction of the potential 
is controlled through  the  space exchange mixture parameter $u$. 
Note, that due to the simple harmonic oscillator shell 
model description of the alpha particle, the change of the potential parameter 
$u$ does not change the energy of the $\alpha$-particle and thus the 
three-body threshold remains the same. The wave function of the system
is taken as linear combination of terms describing the $\alpha(NN)$ and 
$(\alpha N)N$ arrangements. The details of the model is given elsewhere
\cite{asv}. 
To compare the results to other calculations the parameters and the
model space of Ref.\cite{csoto} have been adopted. 

\par\indent
Table I compares the results of ACCC with those of CSM. The CSM is not an
exact solution, and especially for complex systems, the results of CSM might 
have inherited some inaccuracies 
from the underlying gaussian expansion and numerical inaccuracies due to 
the complex arithmetics \cite{csoto}. The results of ACCC and CSM therefore
should be consistent within ``error bars''. Bearing in mind that both methods 
attempt to solve a composite system starting from a fully microscopic model, 
the agreement of the result can be considered to be very good. 
These examples show that the ACCC can be combined with a microscopic structure 
model. The ACCC might be applied in combination with other microscopic 
method like shell model or Hartree-Fock method. 
\par\indent
In a three-body system, the form of singularity in the coupling constant 
is likely to be different from that of the two-body case. The need for higher
order terms in the Pade-approximation may reflect this incorrectness.  
The knowledge of the analytical form of the near threshold singularities would
make the convergence (in the terms of Pade-expansion) faster.
\par\indent
In summary, we have shown through various  
examples that the ACCC is really a powerful method to cope with resonances
of nuclear systems. The application of the ACCC method has been made possible
by the recent developments in solving bound state type problems and the 
increase of computational power. The results of the ACCC method have been 
compared to those of other solutions and are found to be in good agreement.
\par\indent
The advantage of the ACCC over the CSM is that one does not need to calculate
complex matrix elements, that one does not have to use complex arithmetics 
on the computer. Unlike the real stabilization method, one does not need to 
recalculate the Hamiltonian on different bases many times. On the other hand,
to have a reliable solution by the ACCC method, one has to solve the bound 
state problems very accurately. This requirement may pose certain limitations 
in applications. The determination of resonances by the direct solution of 
the Faddeev-equation \cite{sop1,sop2} is certainly superior, but the ACCC
seems to be more easily applicable in the framework of microscopic models.
\par\indent
The results encourage the application of ACCC to study the resonance 
states of few-nucleon and few-cluster systems.
\par\indent
The authors greatly acknowledge the discussions by Prof. V. I. Kukulin and
Dr. A. Cs\'ot\'o. 
This work was supported by Grants-in-Aid for Scientific Research 
( No. 05243102 and No. 06640381 ) and for International Scientific Research
( Joint Research )( No. 08044065 ) of the Ministry of Education, Science 
and Culture (Japan) and by OTKA grant No.T17298 (Hungary). 
K.V. is grateful to the support of Japan Society for the Promotion of Science.
%%%%%%%%%%%%

\begin{figure}[h]
\epsfxsize=8.5cm\epsffile{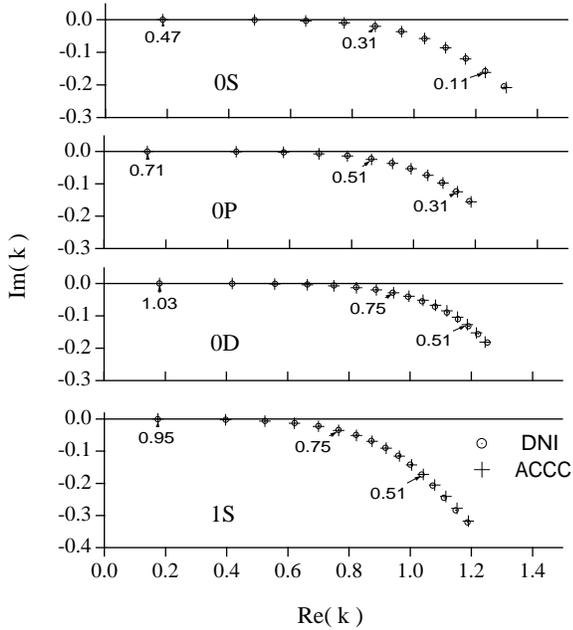}
\caption{The resonances for the two-range Gaussian potential 
of Eq.(2). The ACCC results with $M$=$N$=5 are compared with those of 
direct numerical integration (DNI). The trajectories of the resonances are 
plotted as a function of the coupling constant $\lambda$ 
at intervals of $\delta \lambda = 0.04$.}
\end{figure}

\begin{figure}[h]
\epsfxsize=8.5cm\epsffile{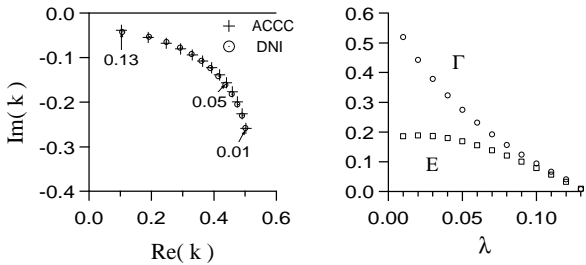}
\caption{The trajectories of wide {\it S}-wave resonances. 
The same as Fig.1 but the barrier height is reduced as described in text. 
The Pad\'e approximation with $M$=$N$=19 is used. 
The energy and the width of the resonance are also shown as a function 
of $\lambda$.}
\end{figure}

\begin{figure}[h]
\epsfxsize=8.5cm\epsffile{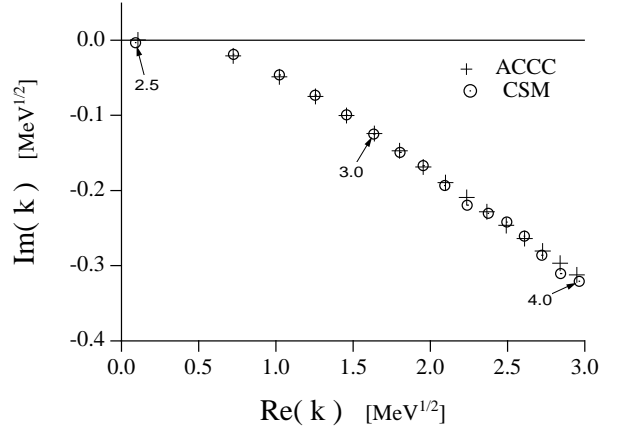}
\caption{The resonances of the three-boson system 
interacting via the potential of Eq. (3). The ACCC results with $M$=$N$=7 
are compared with those of the complex scaling method (CSM). 
The trajectories of the resonances are plotted as a function 
of the coupling constant $\lambda$ at intervals of $\delta \lambda = 0.1$.}
\end{figure}

\begin{center}
\begin{table}
\caption{Comparison of resonance energy and width between ACCC and CSM. The energy is from the three-body threshold.}
\begin{tabular}{lll}
$^{6}$He($2^{+}$,T=1) ($M,N$) & $E$ [MeV] & $\Gamma$ [MeV] \\ \hline
\hspace{5mm}ACCC \hspace{5mm}(9,9)& 0.73      & 0.07         \\
\hspace{5mm}CSM$^{\rm a)}$ & 0.74      & 0.06         \\
\hspace{5mm}Exp$^{\rm b)}$ & 0.82 $\pm$ 0.025 & 0.133 $\pm$ 0.020 \\ \hline
$^{6}$Li($0^{+}$,T=1)  & & \\ \hline
\hspace{5mm}ACCC \hspace{5mm}(9,9)& 0.21      & 0.003        \\
\hspace{5mm}CSM$^{\rm a)}$ & 0.22      & 0.001        \\
\hspace{5mm}Exp$^{\rm b)}$ & $-$0.137  &   \\ \hline
$^{6}$Li($2^{+}$,T=1)  & &  \\ \hline
\hspace{5mm}ACCC \hspace{5mm}(9,9)& 1.61      & 0.27         \\
\hspace{5mm}CSM$^{\rm a)}$ & 1.59      & 0.28         \\
\hspace{5mm}Exp$^{\rm b)}$ & 1.696 $\pm$ 0.015 & 0.54 $\pm$ 0.020 \\
\end{tabular}
\begin{flushleft}
a) Ref \cite{csoto}, \ b) Ref \cite{exp}
\end{flushleft}
\end{table}
\end{center}

\end{document}